\documentstyle[11pt,newpasp,twoside,epsf]{article}
\begin{document}
\title{UBVRI photometric sequences for symbiotic stars}
 \author{Ulisse Munari}
\affil{Osservatorio Astronomico di Padova -- INAF, Sede di Asiago, I-36012 Asiago (VI), Italy}
 \author{Arne Henden}
\affil{Universities Space Research Association/U.S. Naval Observatory Flagstaff Station,
P.O. Box 1149, Flagstaff AZ 86002-1149, USA}

\begin{abstract}
Deep, wide-range and accurate UBVRI photometric sequences have been established
around more than 80 symbiotic stars, to assist current photometry as well
as measurement of old photographic plates. Sequences for 40 symbiotic stars have
already been published; the observations for the others have already
been secured.
\end{abstract}

\section{Introduction}

Availability of suitable UBVRI photometric sequences around symbiotic stars
is an essential step in promoting a large-scale documentation effort of
present time multi-band lightcurves as well as reconstruction of the
historical behavior.

To serve for measuring archival plates, the sequence stars have to be well
isolated from neighbors to avoid blending on short focal length patrol
plates, have to be grouped close to the symbiotic variable so as to
fall within the same eyepiece field of view, and must extend over
a wide range in brightness to easily cover both faint states in
quiescence (like eclipses) as well as bright outbursts.

To assist present time CCD photometry (carried out mostly by amateurs with
telescopes of 1-2 m focal lengths), the sequences must also cover a
wide range in colors (to allow simultaneous derivation of transformation
coefficients), be compact enough to easily fit onto a small CCD ($\la$5 arcmin
in diameter) and be accurately placed on the UBVRI system.

The UBVRI sequences on the Johnson-Cousins system that we have calibrated
around more than 80 symbiotic stars follow the above prescriptions and are
discussed in this paper. Their use on Asiago archive plates is described by
Jurdana-Sepic and Munari (2002, and references therein).

\section{The sequences}

Figure~1 and Table~1 show a typical UBVRI photometric sequence among
the 82 we have calibrated. The increasing errors at the brightest magnitudes
are due to saturation effects, and those at the
faintest magnitudes are dominated by the photon statistics.

\begin{figure}[!t]
\plotone{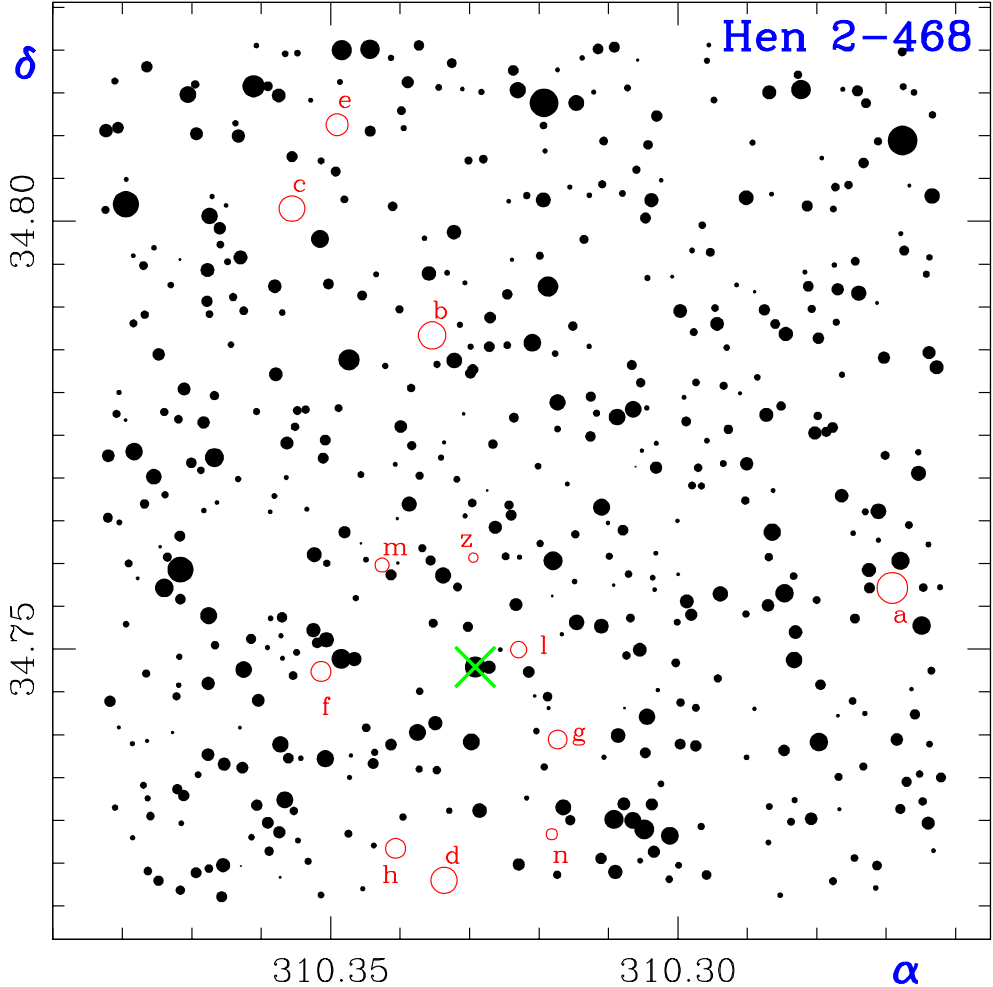}
\caption{Identification chart for the UBVRI sequence in Table~1.}
\end{figure}

The symbiotic stars for which we have already obtained the multi-epoch UBVRI
observations suitable for the calibration of the photometric sequences are
listed in Table~2. Of the 82 listed objects, 40 have their sequences 
published by Henden and Munari (2000, 2001); the remaining 42 will follow
shortly. The program stars have been mainly selected from the catalogues of
symbiotic stars of Allen (1984) and Belczy\'nski et al. (2000)

\begin{table}[!t]
\begin{center}
\plotone{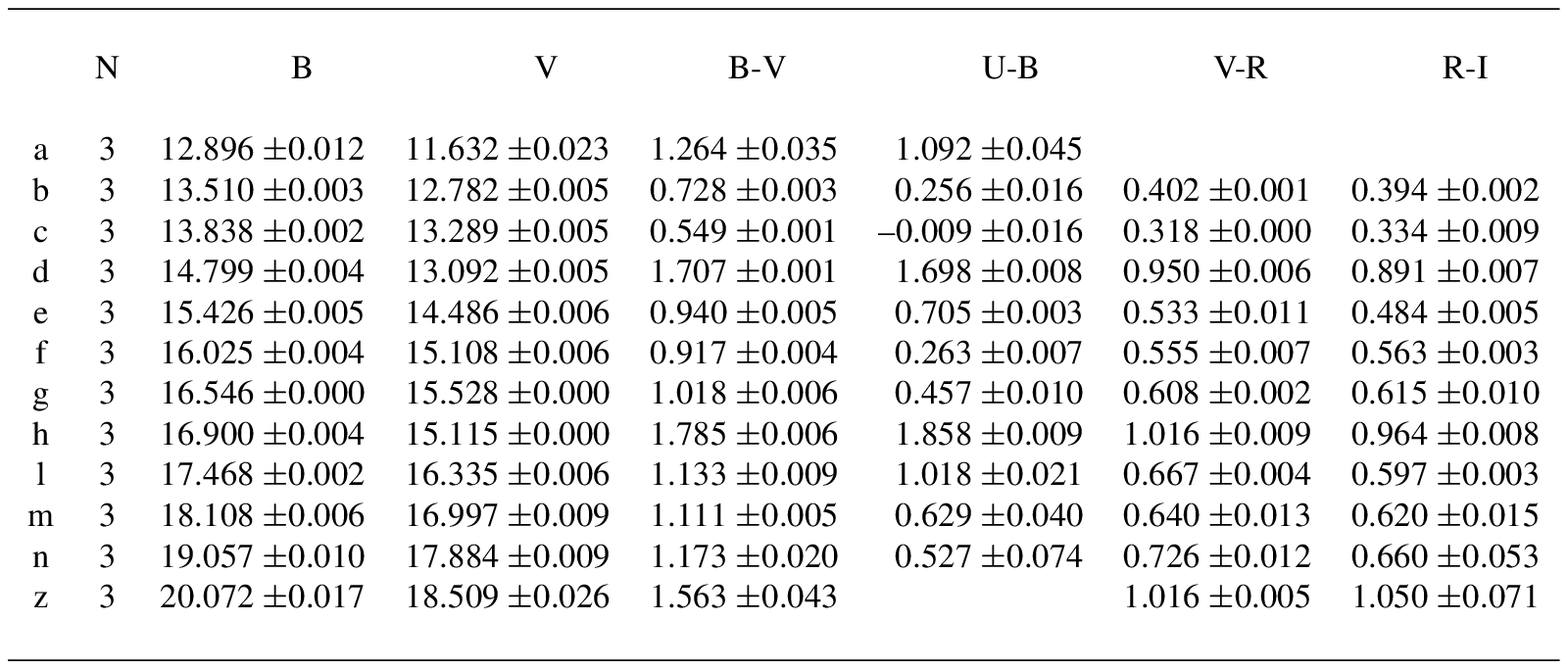}
\caption{The UBVRI comparison sequence plotted in Figure~1. $N$ is
the number of observations on separate nights used to check for
variability. The quoted errors are formal ones.}
\end{center}
\end{table}

\begin{table}[!b]
\plotone{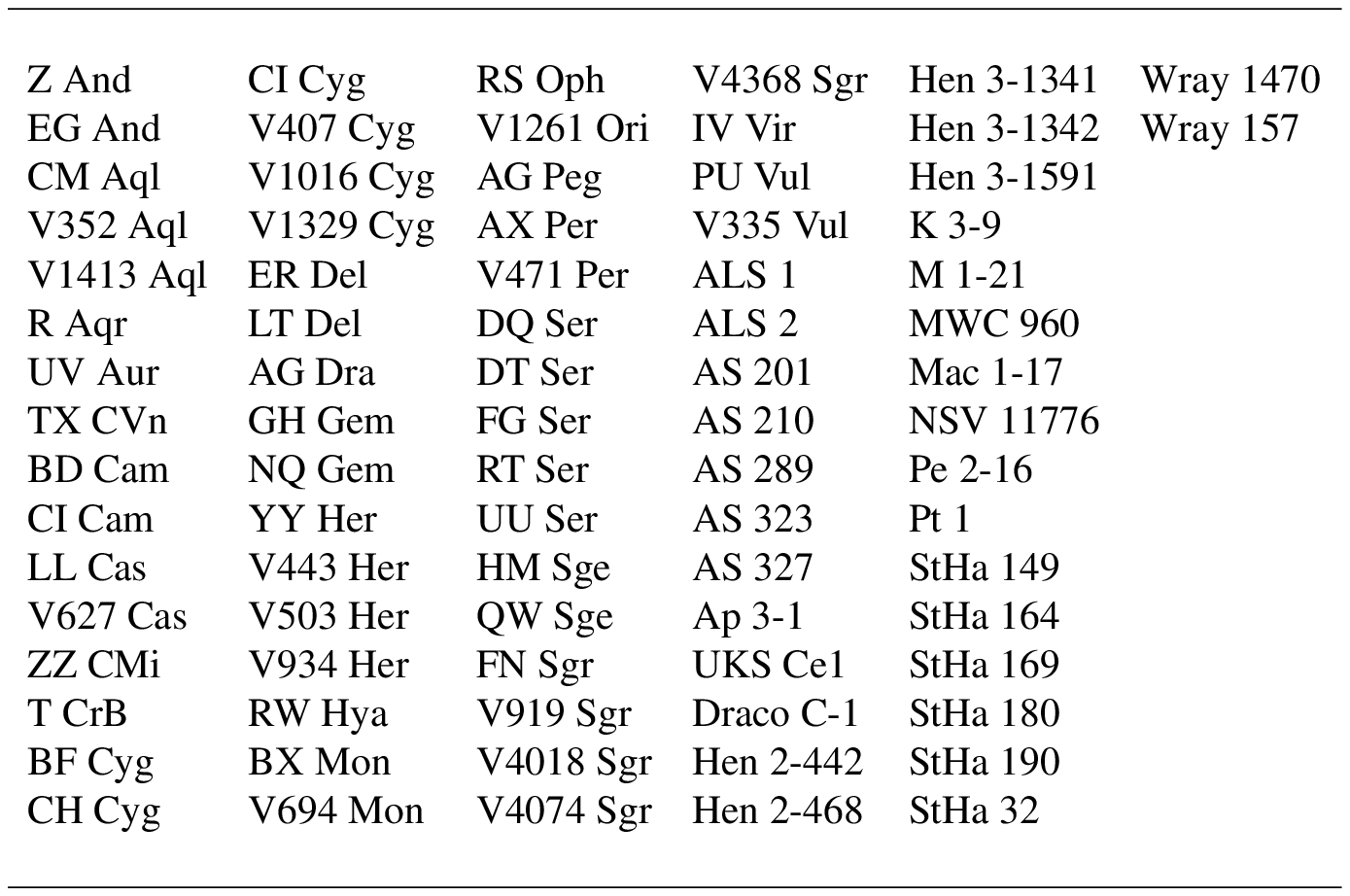}
\caption{List of symbiotic stars for which we have obtained a UBVRI photometric
sequence. Forty of them have been published in Henden and Munari (2000, 2001);
for the remaining 42, the observations have been completed and
publication will follow.}
\end{table}

All observations were made with the 1.0 m Ritchey-Chr\'etien telescope of
the U.S. Naval Observatory, Flagstaff Station. A Tektronix/SITe
1024$\times$1024 thinned, backside-illuminated CCD was used, along with
proper Johnson UBV and Kron-Cousins RI filters. The telescope scale is 0.67
arcsec/pix and typical seeing was $\sim$2 arcsec. The color transformation
equations for this instrumental combination are particularly good, with
slopes deviating from unity by only a few hundredths of a magnitude (see
Eq.(1)-(5) in Henden and Munari 2000). Only data collected under
photometric conditions (transformation errors under 0.02 mag) have
been used, and the calibration has been made against the Landolt
equatorial fields (Landolt 1983, 1992).

Each photometric sequence is composed of 10 to 15 stars,
which have been selected and ordered on the basis of their $B$ band
magnitude. The latter is reproducible by most filter-equipped CCD cameras,
it is the closest one to the $m_{pg}$ band of historical observations, and
the $B$ band is particularly well suited to investigate the variability of
symbiotic stars. Stars of varying color are included in the sequence to
allow simultaneous determination of transformation coefficients.
The comparison sequences are tighter around the usual brightness of
each symbiotic star and become looser away from it.

The stars included in the comparison sequences have been checked on at least
three nights for variability. We cannot obviously rule out beyond doubt that
some of them are indeed variable, but the agreement at a few milli-mag level
of their magnitude measured on different nights over an interval of months
gives some confidence in their use.

\end{document}